\documentclass[11pt, preprint]{revtex4}

\usepackage{graphicx}
\usepackage{bm}

\begin{document}

\title{Turing patterns in network-organized activator-inhibitor systems}

\begin{abstract}
  Turing instability in activator-inhibitor systems provides a
  paradigm of nonequilibrium pattern formation; it has been
  extensively investigated for biological and chemical processes.
  Turing pattern formation should furthermore be possible in
  network-organized systems, such as cellular networks in
  morphogenesis and ecological metapopulations with dispersal
  connections between habitats, but investigations have so far been
  restricted to regular lattices and small networks.  Here we report
  the first systematic investigation of Turing patterns in large
  random networks, which reveals their striking difference from the
  known classical behavior.  In such networks, Turing instability
  leads to spontaneous differentiation of the network nodes into
  activator-rich and activator-low groups, but ordered periodic
  structures never develop.  Only a subset of nodes having close
  degrees (numbers of links) undergoes differentiation, with its
  characteristic degree obeying a simple general law.  Strong
  nonlinear restructuring process leads to multiple coexisting states
  and hysteresis effects.  The final stationary patterns can be well
  understood in the framework of the mean-field approximation for
  network dynamics.
\end{abstract}

\author{Hiroya Nakao$^{*}$}

\address{Department~of~Physics, Kyoto~University, Kyoto~606-8502,
  Japan\\
  Phone:~+81-75-753-3742~~FAX:~+81-75-753-3819\\
  E-mail: nakao@ton.scphys.kyoto-u.ac.jp}

\author{Alexander S. Mikhailov}

\address{Department~of~Physical~Chemistry,
  Fritz~Haber~Institute~of~the~Max~Planck~Society, Faradayweg 4-6, 14195 Berlin, Germany\\
  Phone:~+49-30-8413-5122~~FAX: +49-30-8413-5106\\
  E-mail: mikhailov@fhi-berlin.mpg.de}

\date{\today}

\maketitle

\section{Introduction}

In 1952, A. Turing published a seminal paper~\cite{Turing},
demonstrating that differences in diffusion rates of reacting species
can alone destabilize the uniform state of the system and lead to the
formation of spatial patterns and suggesting this as a possible
mechanism of biological morphogenesis.
The Turing instability and resulting patterns have subsequently been
theoretically analyzed and experimentally confirmed for various
chemical, biological, and ecological
systems~\cite{Prigogine,Mimura,DeKepper,Ouyang,Sick}; they are
generally viewed as a paradigm of nonequilibrium pattern formation.
In 1971, Othmer and Scriven~\cite{Othmer} pointed out that the Turing
instability is also possible in network-organized systems and this
should be important for the understanding of multi-cellular
morphogenesis. At an early stage of the organism development, a
network of inter-cellular connections is formed and morphogens are
diffusively transported over such a network; differentiation of cells
may thus be induced by the network Turing instability.
On the other hand, many ecosystems represent metapopulations
distributed over discrete habitats forming networks with dispersal
connections~\cite{Hanski,Urban}.  Prey and predator or host and
parasite species may migrate over such networks with different
diffusional mobilities.
Recently, diffusional spreading of infectious diseases over airline
transportation networks has attracted much attention~\cite{Hufnagel}.
Epidemic dynamics on the networks can be described by
reaction-diffusion models where infected and susceptible individuals
play the role of interacting species~\cite{Murray,Liu}.
In all such network-organized systems with diffusional transport of
interacting species, the Turing instabilities and resulting nonlinear
patterns can generally be expected.

Although nonlinear dynamics on complex networks is attracting
significant attention, most of the investigations have focused on
synchronization phenomena of oscillator networks (see, e.g., recent
reviews~\cite{Boccaletti,Arenas}).
Despite the potential importance of network Turing patterns and the
large amount of research on classical Turing patterns in spatially
extended systems, very little research on this problem has been
performed so far.
Early studies by Othmer and Scriven~\cite{Othmer,Othmer2} have
provided abstract mathematical framework for the analysis of network
Turing instability, but they explicitly considered only simple
examples of regular lattices close in their properties to continuous
media.  Recently, Horsthemke {\it et. al.}~\cite{Horsthemke,Moore}
have discussed the possibility of Turing instability in coupled
chemical reactors, but only for small networks.

In this article, we present the systematic analytical and numerical
study of the Turing instability and of the developing nonlinear
patterns in large random networks.  We find that the Turing
instability generally occurs in network-organized activator-inhibitor
systems, but its properties are very different from those
characteristic for the classical continuous media.

\section{Network-organized activator-inhibitor systems}

Classical activator-inhibitor systems in continuous media are
described by equations of the form
\begin{eqnarray}
  \frac{\partial}{\partial t} u({\bm x},t) &=& f(u, v) + D_{act}
  \nabla^2 u({\bm x}, t), \cr \cr
  \frac{\partial}{\partial t} v({\bm x},t) &=& g(u, v) + D_{inh}
  \nabla^2 v({\bm x}, t),
  \label{Eq:CRD}
\end{eqnarray}
where $u({\bm x}, t)$ and $v({\bm x}, t)$ are local densities of the
activator and inhibitor species.  Here, the functions $f(u, v)$ and
$g(u, v)$ specify dynamics of the activator that autocatalytically
enhances its own production and of the inhibitor that suppresses the
activator growth.
$D_{act}$ and $D_{inh}$ are the diffusion constants of the activator
and inhibitor species.  The classical Turing instability~\cite{Turing}
sets in as the ratio $D_{inh} / D_{act}$ of the two diffusion
constants is increased and exceeds a threshold.  It leads to
spontaneous development of alternating activator-rich and
activator-low domains from the uniform background.

In our study, we consider the network analog of the
model~(\ref{Eq:CRD}), where activator and inhibitor species occupy
discrete nodes of a network and are diffusively transported over the
links connecting them.
We consider a connected network consisting of $N$ nodes, $i=1, \cdots,
N$.  The network topology is defined by a symmetric adjacency matrix
whose elements $A_{ij}$ take values $A_{ij} = 1$ if the nodes $i$ and
$j$ are connected $(i \neq j)$ and $A_{ij} = 0$ otherwise.
Diffusive transport of the species into a certain node $i$ is simply
given by the sum of incoming fluxes to the node $i$ from other
connected nodes $\{j\}$, where the fluxes are proportional to the
concentration difference between the nodes (Fick's law).
By introducing the {\em network Laplacian matrix} whose elements are
given by $L_{ij} = A_{ij} - k_i \delta_{ij}$, where $k_i =
\sum_{j=1}^{N} A_{ij}$ is the degree of the node $i$, diffusive flux
of the species $u$ to node $i$ is expressed as $ \sum_{j=1}^{N} L_{ij}
u_j = \sum_{j=1}^{N} A_{ij} (u_j - u_i) $, and similarly for $v$.
Generally, diffusional mobilities of species $u$ and $v$ on the
network are different.
The equations describing network-organized activator-inhibitor
systems are thus given by
\begin{eqnarray}
  \frac{d}{dt} u_i(t) &=& f(u_i, v_i) + \varepsilon \sum_{j=1}^{N} L_{ij} u_j, \cr \cr
  \frac{d}{dt} v_i(t) &=& g(u_i, v_i) + \sigma \varepsilon \sum_{j=1}^{N} L_{ij} v_j,
  \label{Eq:RD}
\end{eqnarray}
for $i=1, \cdots, N$.  Here, $f(u, v)$ and $g(u, v)$ specify the local
activator-inhibitor dynamics on individual nodes.  We denote the
diffusional mobility of the activator species as $\varepsilon
(=D_{act})$ and of the inhibitor species as $\sigma \varepsilon
(=D_{inh})$, so that $\sigma = D_{inh} / D_{act}$ is the ratio between
them.  The considered systems have a uniform stationary state
$(\bar{u}, \bar{v})$, where $f(\bar{u}, \bar{v}) = 0$ and $g(\bar{u},
\bar{v}) = 0$.  This uniform state can become unstable as a result of
the Turing instability. If $u$ and $v$ correspond to the activator and
inhibitor species, functions $f$ and $g$ should satisfy several
conditions which are given in the Methods section.

As the examples of activator-inhibitor systems, we use the
Mimura-Murray model of prey-predator populations~\cite{Mimura} and the
classical Brusselator model~\cite{Prigogine} which are described in
the Methods section.  This study is focused on the Turing instability
and pattern formation in large random networks.  We use scale-free
networks which are ubiquitous in
Nature~\cite{Barabasi,Barabasi2,Boccaletti,Arenas} and the classical
Erd\"os-R\'enyi random networks~\cite{Barabasi,Barabasi2}, both
described in the Methods section.  For convenience, network nodes
$\{i\}$ are always sorted below in the decreasing order of their
degrees $\{k_i\}$, so that the condition $k_1 \geq k_2 \geq \cdots
k_N$ holds.

\section{The Turing instability}

The Turing instability is revealed through the linear stability
analysis of the uniform stationary state with respect to nonuniform
perturbations (see Methods for the details).  In the classical case of
continuous media~\cite{Turing}, nonuniform perturbations are
decomposed over a set of spatial Fourier modes, representing plane
waves with different wavenumbers.  As has been originally noticed by
Othmer and Scriven~\cite{Othmer}, in the networks, the role of plane
waves should be played by eigenvectors of their Laplacian matrices.
The eigenvalues $\Lambda_{\alpha}$ and eigenvectors ${\bm
  \phi}^{(\alpha)} = (\phi_{1}^{(\alpha)}, \cdots,
\phi_{N}^{(\alpha)})$ of the Laplacian matrix $L_{ij}$ are
determined~\cite{Boccaletti,Mohar} by $ \sum_{j=1}^{N} L_{ij}
\phi_{j}^{(\alpha)} = \Lambda_{\alpha}\phi_{i}^{(\alpha)} $, with
$\alpha= 1, \cdots, N$.  All eigenvalues of $L_{ij}$ are non-positive.
We sort the indices $\{\alpha\}$ in the decreasing order of the
eigenvalues, so that the condition $0 = \Lambda_{1} \geq \Lambda_{2}
\geq \cdots \geq \Lambda_{N}$ always holds.

Introducing small perturbations $( \delta u_i, \delta v_i )$ to the
uniform state as $ ( u_{i}, v_{i} ) = ( \bar{u}, \bar{v} ) + ( \delta
u_i, \delta v_i )$ and substituting this into equations~(\ref{Eq:RD}), a
set of coupled linearized differential equations is obtained.  By
expanding the perturbations over the set of Laplacian eigenvectors as
$\delta u_i(t) = \sum_{\alpha=1}^{N} c_{\alpha} \exp\left[
  \lambda_{\alpha} t \right] \phi_i^{(\alpha)}$ and $\delta v_i(t) =
\sum_{\alpha=1}^{N} c_{\alpha} B_{\alpha} \exp\left[ \lambda_{\alpha}
  t \right] \phi_i^{(\alpha)}$, these equations are transformed into
$N$ independent linear equations for different normal modes.  The
linear growth rate $\lambda_{\alpha}$ of the $\alpha$-th mode is
determined from the characteristic equation.
When $\mbox{Re}\ \lambda_{\alpha}$ is positive, the $\alpha$-th mode
is unstable. The Turing instability takes place when one of the modes
(i.e., {\em the critical mode}) begins to grow.  At the instability
threshold, $\mbox{Re }\lambda_{\alpha} = 0$ for some $\alpha =
\alpha_{c}$ and $\mbox{Re }\lambda_{\alpha} < 0$ for all other modes.
In the Turing instability, the critical mode is not oscillatory,
$\mbox{Im}\ \lambda_{\alpha_c}$ = 0.

\begin{figure}[htbp]
  \begin{center}
    \includegraphics[width=1.0\hsize]{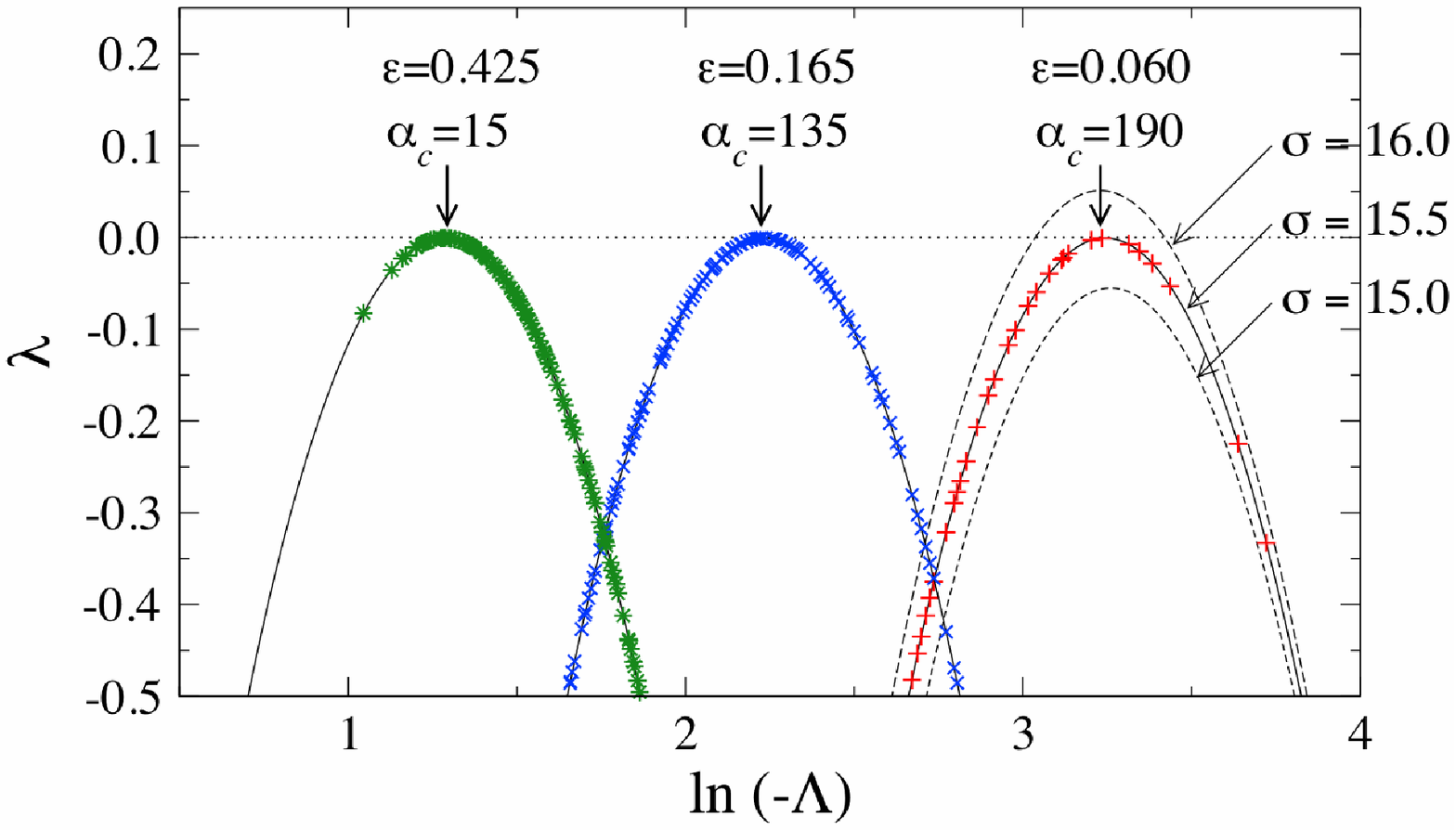}
  \end{center}
  \caption{Linear stability analysis.  Linear growth rates
    $\lambda_{\alpha}$ of the Laplacian modes $\alpha = 1, \cdots, N$
    for the Mimura-Murray model on a scale-free network ($N=200$ nodes
    and mean degree $\langle k \rangle = 10$) for three values of the
    diffusional mobility $\varepsilon$ and for the critical ratio of
    diffusion constants $\sigma = 15.5 \simeq \sigma_c$.  Three curves
    corresponding $\varepsilon = 0.425$, $0.165$, and $0.060$ are
    plotted as functions of the Laplacian eigenvalues
    $\Lambda_{\alpha}$.  For comparison, curves with $\sigma = 15.0$
    and $\sigma = 16.0$ are also drawn for $\varepsilon = 0.060$.
    Critical modes are indicated for each value of $\varepsilon$.  The
    critical modes and the corresponding Laplacian eigenvalues are
    $\alpha_c = 15$, $\Lambda_{c} = -3.62$ for $\varepsilon = 0.425$,
    $\alpha_c = 135$, $\Lambda_{c} = -9.32$ for $\varepsilon = 0.165$,
    and $\alpha_c = 190$, $\Lambda_{c} = -25.3$ for $\varepsilon =
    0.060$.}
  \label{Fig1}
\end{figure}

As an example, Fig.~\ref{Fig1} shows the growth rate $\lambda$ as a
function of $\Lambda$ for the Mimura-Murray model. At $\varepsilon =
0.060$, three curves corresponding to different ratios $\sigma$ of
diffusion constants (below, at and above the instability threshold)
are displayed. In this figure, critical curves for two other values of
the parameter $\varepsilon$ are also shown.

Generally, the Turing instability becomes possible for $\sigma >
\sigma_c$.  The dispersion curve $\lambda = F(\varepsilon \Lambda)$ first
touches the horizontal axis at $\Lambda = \Lambda_c$ and the
Laplacian mode $\phi_i^{(\alpha_c)}$, possessing the Laplacian
eigenvalue $\Lambda_{\alpha_c}$ that is closest to $\Lambda_c$,
becomes critical.  For the critical mode, the coefficient
$B_{\alpha}$ is positive, so that when the activator concentration
increases, the inhibitor concentration also increases accordingly.
Explicit expressions are given in the Methods section.
Note that the Laplacian spectrum of
a network is discrete and, therefore, the instability actually occurs
only when one of the respective points on the dispersion curve crosses
the horizontal axis.

The above results are analogous to those holding for continuous media
(cf.~\cite{Mikhailov}). The critical ratio $\sigma_c$ in the networks
is the same as in the classical case. The Laplacian eigenvalue
$\Lambda_c$ of the critical network mode corresponds to $-q_c^2$,
where $q_c$ is the wavenumber of the critical mode in the continuous
media.  Despite such formal analogies, properties of the network
Turing patterns are very much different from their classical
counterparts, as demonstrated in the following sections.

\section{Localization of Laplacian eigenvectors and characteristic
  properties of critical Turing modes}

When a Turing pattern starts to grow after slightly exceeding the
instability threshold, the activator distribution in this pattern is
determined by the critical Laplacian eigenvector, i.e. we have $\delta
u_i \propto \phi_i^{(\alpha_c)}$. Therefore, to understand
organization of the growing Turing patterns, properties of Laplacian
eigenvectors should be considered.

\begin{figure}[htbp]
  \begin{center}
    \includegraphics[width=1.0\hsize]{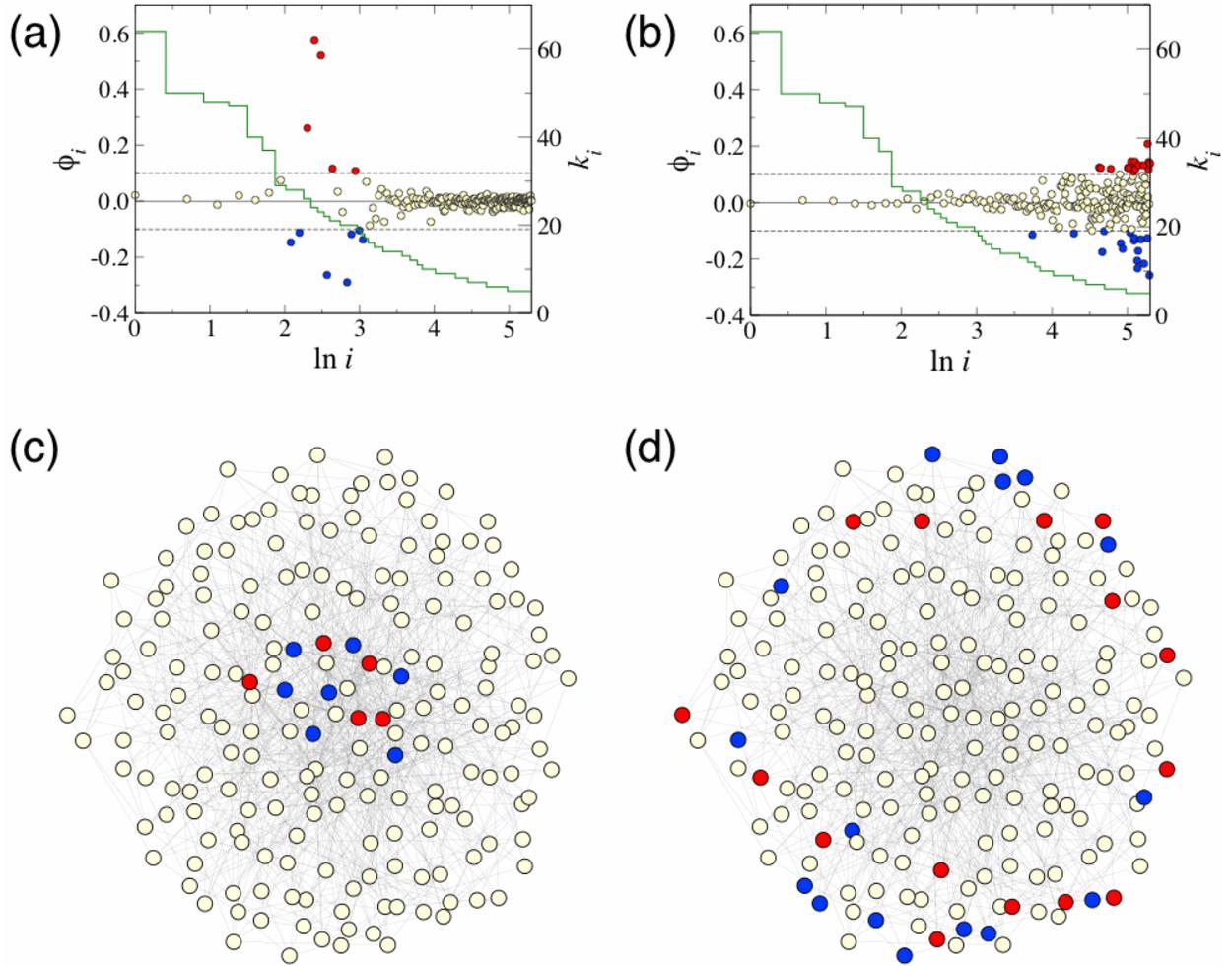}
  \end{center}
  \caption{Critical Turing modes for a scale-free network of size
    $N=200$ and mean degree $\langle k \rangle = 10$.  (a,b): Critical
    eigenvectors (a) $\alpha_c = 190$ and (b) $\alpha_c = 15$ plotted
    against the node index $i$.  Node degrees $k_i$ are shown by green
    stepwise curves.  Node indices $\{i\}$ are sorted according to
    their degrees $\{k_i\}$.  (c,d): The same critical eigenvectors
    (c) $\alpha_c = 190$ and (d) $\alpha_c = 15$ displayed graphically
    on the network.}
  \label{Fig2}
\end{figure}

As an example, Figs.~\ref{Fig2}(a,b) display critical eigenvectors of
a network for two different values of the diffusion constant
$\varepsilon$. The same eigenvectors are shown graphically in
Figs.~\ref{Fig2}(c,d).  In the chosen representation, network nodes
with larger degrees (hubs) are located in the center and the nodes
with lower degrees in the periphery of the graph.  The nodes are
colored red when $\phi_i^{(\alpha_c)} \geq 0.1$ (e.g. activator
concentration is significantly increased), blue when
$\phi_i^{(\alpha_c)} \leq -0.1$ (significantly decreased), and yellow
for $-0.1 < \phi_{i}^{(\alpha_c)} < 0.1$ (no significant change).

It is clearly seen in Fig.~\ref{Fig2} that spontaneous differentiation
of the nodes takes place - the distinguishing feature of the Turing
instability.  However, it affects only a fraction of all nodes.  The
differentiated nodes, with significant deviations of the activation
level, tend to have close degrees.  When the diffusional mobility
$\varepsilon$ is small, only a subset of hub nodes undergoes
differentiation [Figs.~\ref{Fig2}(a),(c)]. If $\varepsilon$ is large,
differentiated nodes have just a few links [Figs.~\ref{Fig2}(b),(d)].
Thus, a correlation between the characteristic degrees of the
differentiated nodes and the diffusional mobility is present.  The
behavior observed in Fig.~\ref{Fig2} is general. As we show below, it
is related to the effect of localization of Laplacian eigenvectors in
large random networks.

\begin{figure}[htbp]
  \begin{center}
    \includegraphics[width=1.0\hsize]{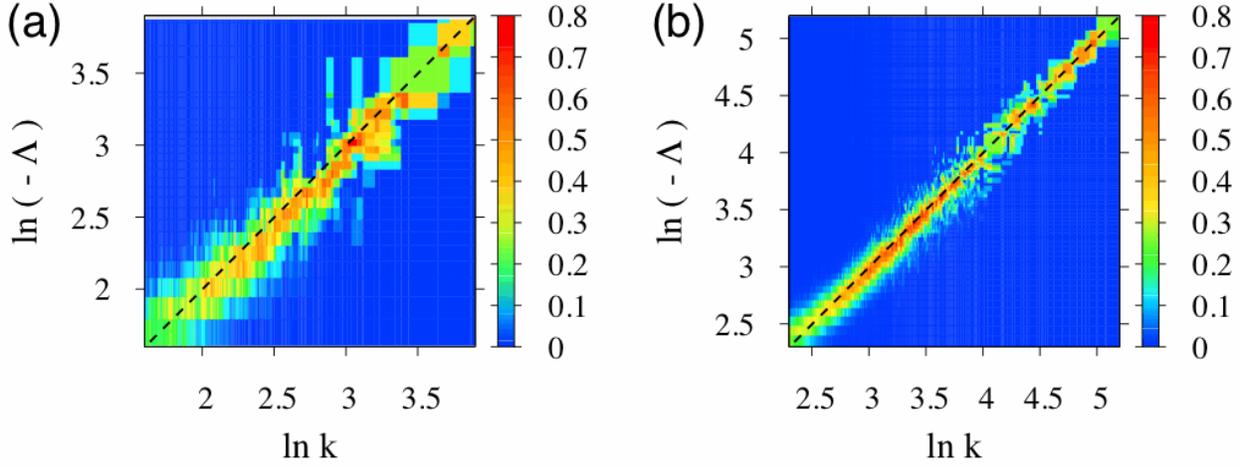}
  \end{center}
  \caption{Localization of Laplacian eigenvectors in scale-free
    networks.  The network size and mean degree are (a) $N=200$,
    $\langle k \rangle = 10$ and (b) $N=1000$, $\langle k \rangle =
    20$.  All nodes are divided into groups with equal degrees.  For
    each group, the number of ``differentiated'' nodes with
    $\phi_{i}^{(\alpha)} \geq 0.1$ or $\phi_{i}^{(\alpha)} \leq -0.1$
    for each eigenvector $\alpha$ is counted.  Then the fraction of
    such nodes in each group for each Laplacian eigenvector $\alpha$
    is determined.  Thus, these diagrams show density distributions of
    differentiated nodes for the entire set of Laplacian
    eigenvectors.}
  \label{Fig3}
\end{figure}

As has recently been discussed~\cite{Menzinger}, Laplacian
eigenvectors in large random networks with relatively broad degree
distributions tend to localize on subsets of nodes with close degrees.
The localization effect for a scale-free network is illustrated in
Fig.~\ref{Fig3}.  Here, all nodes are divided into groups with equal
degrees $k$. For each group $k$ and a given Laplacian eigenvalue
$\Lambda$, the number of ``differentiated'' nodes with
$\phi_i^{(\alpha)} \geq 0.1$ or $\phi_i^{(\alpha)} \leq -0.1$ in the
respective eigenvector $\phi_i^{(\alpha)}$ is counted. The density
diagrams in Fig. 3 display in the color code the relative numbers of
such nodes as functions of the Laplacian eigenvalue $\Lambda$ and the
degree $k$.  Examining Fig.~\ref{Fig3}, one can see that differentiated nodes
are approximately located along the diagonal of the density map.  The
localization effect is more pronounced for the larger network of size
$N = 1000$. Similar localization behavior is observed for the
Erd\"os-R\'enyi networks (see Supplementary information). Thus, we see
that each Laplacian eigenvector $\phi_i^{(\alpha)}$ is characterized
by its characteristic localization degree
$\bar{k}_{\alpha}$. Moreover, this characteristic degree is
approximately equal to the negative of the respective eigenvalue,
so that a simple relationship
$\bar{k}_{\alpha} \simeq - \Lambda_{\alpha}$ holds.

On the other hand, as implied by the linear stability analysis (see
Methods), the growth rate $\lambda_{\alpha}$ of each mode depends only
on the combination $\varepsilon \Lambda_{\alpha}$ of the diffusional
mobility $\varepsilon$ and the eigenvalue $\Lambda_{\alpha}$ of that mode,
i.e. we have $\lambda_{\alpha} = F(\varepsilon \Lambda_{\alpha})$.
Therefore, the Laplacian eigenvalue $\Lambda_{\alpha_c}$ of the
critical mode $\alpha_c$ with $\lambda_{\alpha_{c}}=0$ should be
inversely proportional to the diffusive mobility $\varepsilon$,
i.e., $\Lambda_{\alpha_c} \propto 1 / \varepsilon$.  Hence, modes
with large negative eigenvalues $\Lambda_{\alpha}$ tend to become
unstable for the small mobilities
$\varepsilon$ (note that $\Lambda_{\alpha} \leq 0$ in our definition).

Combining the two relationships, $\bar{k}_{\alpha} \simeq -
\Lambda_{\alpha}$ and $\Lambda_{\alpha_c} \propto 1 / \varepsilon$, a
simple scaling law $\bar{k}_{\alpha_c} \propto 1/\varepsilon$ is
obtained.  It implies that the characteristic degree
$\bar{k}_{\alpha_c}$ of the differentiating subset is inversely
proportional to the diffusional mobility $\varepsilon$.

The dependence $\Lambda_{\alpha_c} \propto 1/\varepsilon$ holds for
any activator-inhibitor model exhibiting the Turing instability.  The
localization of Laplacian eigenvectors has been observed by us (to be
separately published) also for other random networks with relatively
broad degree distributions.  The characteristic localization degree
$\bar{k}_{\alpha_c}$ of the critical Turing mode is generally a
monotonously increasing function of the negative of the
critical Laplacian eigenvalue, $- \Lambda_{\alpha_c}$, and thus a
decreasing function of the diffusional mobility $\varepsilon$.

\section{Stationary Turing patterns}

\begin{figure}[ptbh]
  \begin{center}
    \includegraphics[width=1.0\hsize]{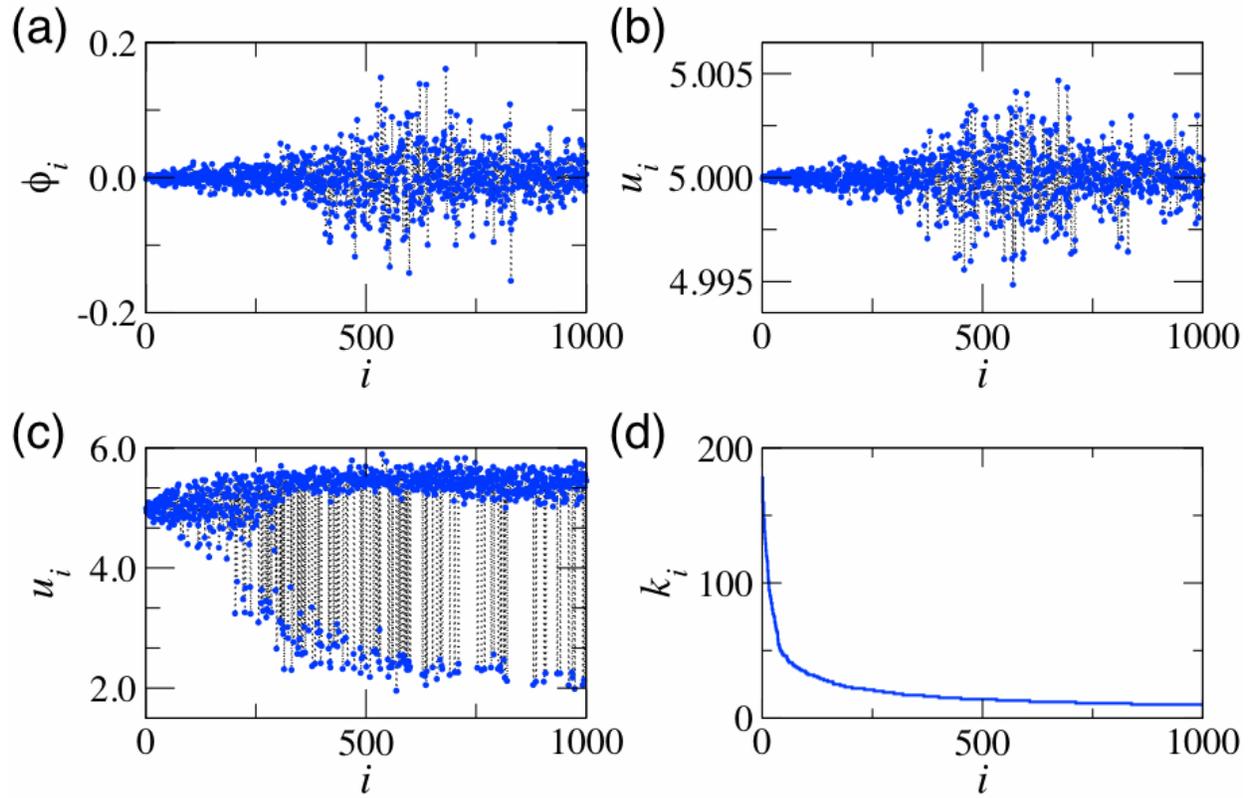}
  \end{center}
  \caption{Nonlinear evolution and a stationary Turing pattern of the
    Mimura-Murray model on a scale-free network at $\epsilon = 0.12$
    and $\sigma = 15.6$.  The network size and mean degree are
    $N=1000$ and $\langle k \rangle = 20$.  (a) The critical mode (the
    Laplacian eigenvector with $\alpha_c = 422$), (b) the activator
    pattern at the early evolution stage ($t = 200$), and (c) the
    stationary activator pattern at the late stage ($t = 1500$).
    Nodes are ordered according to their degrees; with (d) showing the
    dependence of the degree on the node index.}
\label{Fig4}
\end{figure}

The initial exponential growth is followed by a nonlinear process,
leading to the formation of stationary Turing patterns. The nonlinear
evolution of the system and the properties of emerging stationary
patterns have been studied by us in numerical
simulations. Figure~\ref{Fig4} presents typical results, obtained for
intermediate diffusional mobility $(\varepsilon = 0.12)$ and slightly
above the instability threshold $(\sigma = 15.6)$ for the
Mimura-Murray model on the random scale-free network of size $N =
1000$ and mean degree $\langle k \rangle = 20$.  The nodes are sorted
in the order of their degrees, as shown in Fig.~\ref{Fig4}(d).

Starting from almost uniform initial conditions with small
perturbations, exponential growth is observed at the early stage. The
activator pattern at this stage, Fig.~\ref{Fig4}(b), is similar to the
critical mode, Fig.~\ref{Fig4}(a), with the deviations due to the
contributions from neighboring modes that are already excited to some
extent.  Later on, however, strong nonlinear effects develop, and the
final stationary pattern, Fig.~\ref{Fig4}(c), becomes very different
from the one determined by the critical mode.

Observing the nonlinear development, we notice that some nodes get
progressively kicked off the main group near the destabilized uniform
solution in this process (see Video in the Supplementary information).
Eventually, in the asymptotic stationary state, the nodes become
separated into two groups.  The separation into two groups occurs only
for the nodes with relatively small degrees (roughly $i>200$,
$k_{i}<24$), while the nodes with high degrees ($i<200$, $k_{i}>24$)
do not undergo the differentiation.

\begin{figure}[ptbh]
  \begin{center}
    \includegraphics[width=1.0\hsize]{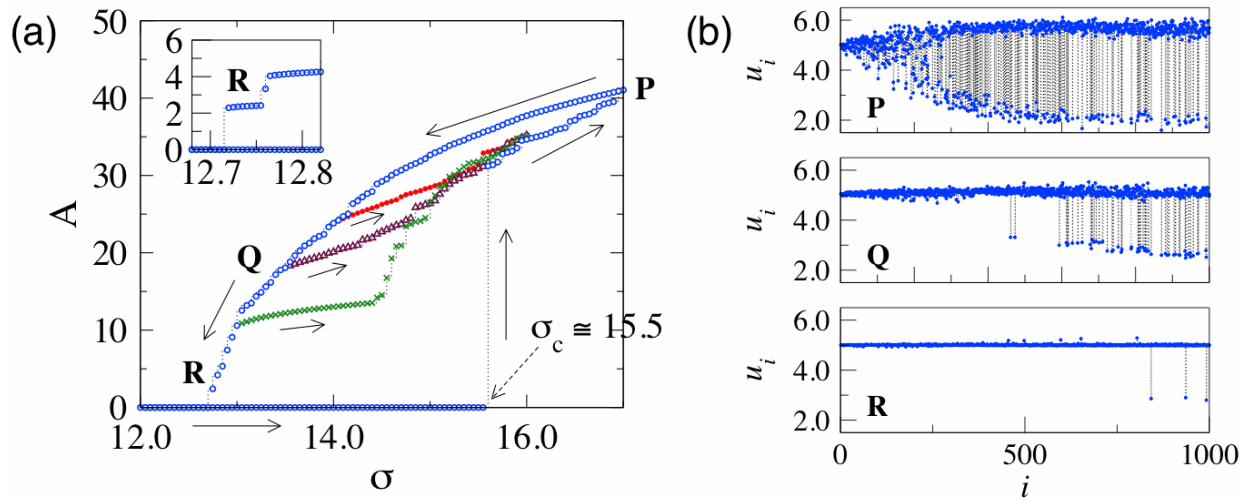}
  \end{center}
  \caption{Hysteresis and multistability.  (a) Amplitude $A$ of the
    Turing pattern vs. the diffusion ratio $\sigma$; variation
    directions of $\sigma$ are indicated by arrows.  The inset shows
    the blowup near $R$.  (b) Stationary Turing patterns at the
    parameter points $P$ ($\sigma=17.0$), $Q$ ($\sigma=13.5$), and $R$
    ($\sigma=12.8$).}
\label{Fig5}
\end{figure}

Our numerical investigations furthermore reveal that the outcome of
nonlinear evolution depends sensitively on initial
conditions. Different Turing patterns are possible at the same
parameter values and strong hysteresis effects are observed. As an
example, Fig.~\ref{Fig5}(a) shows how the amplitude of the stationary
Turing pattern, defined as $A = \left[ \sum_{i=1}^{N}\left\{
    (u_{i}-\bar{u})^{2}+(v_{i}-\bar{v})^{2}\right\} \right]^{1/2}$,
varies under gradual variation of the parameter $\sigma$ in the upward
or downward directions.  Stationary patterns observed at points $P$,
$Q$, and $R$ in Fig.~\ref{Fig5}(a) are shown in Fig.~\ref{Fig5}(b).

As $\sigma$ was increased starting from the uniform initial condition,
the Turing instability took place at $\sigma=\sigma_{c}$, with the
amplitude $A$ suddenly jumping up to a high value that corresponds to
the appearance of a kicked-off group. If $\sigma$ was further
increased, the amplitude $A$ grew. Starting to decrease $\sigma$, we
did not however observe a drop down at $\sigma=\sigma_{c}$. Instead, a
punctuated decrease in the amplitude $A$, which is characterized by
many relatively small steps, was found.  Reversing the direction of
change of the parameter $\sigma$ at different points, many coexisting
solution branches could be identified.  The characteristics of Turing
patterns vary with their amplitudes.  When $A$ is close to zero [
point $R$ in Fig.~\ref{Fig5}(a) ], only a few kicked-off nodes remain
in the system.  Such localized Turing patterns with only a small
number of destabilized nodes can coexist with the linearly stable
uniform state and are found below the Turing instability threshold,
for $\sigma < \sigma_{c}$.

\section{The mean-field theory}

To understand properties of the developed Turing patterns above the
instability boundary ($\sigma > \sigma_{c}$), one can use the
mean-field approximation, similar to that previously employed in the
investigations of epidemics spreading on networks~\cite{Pastor} and
for the networks of phase oscillators~\cite{Ichinomiya}.  We start by
writing equations~(\ref{Eq:RD}) in the form
\begin{eqnarray}
  \frac{d}{dt} u_{i}(t) &=& f(u_{i}, v_{i}) + \varepsilon (h_{i}^{(u)} - k_{i}u_{i}),\cr \cr
  \frac{d}{dt} v_{i}(t) &=& g(u_{i}, v_{i}) + \sigma \varepsilon (h_{i}^{(v)} - k_{i} v_{i}),
\end{eqnarray}
where local fields felt by each node, $h_{i}^{(u)}=\sum_{j=1}^{N}
A_{ij}u_{j}$ and $h_{i}^{(v)}=\sum_{j=1}^{N}A_{ij}v_{j}$, are
introduced.  These local fields are further approximated as
$h_{i}^{(u)}\simeq k_{i} H^{(u)}$ and $h_{i}^{(v)}\simeq
k_{i}H^{(v)}$, where \emph{global mean fields} are defined by
$H^{(u)}=\sum_{j=1}^{N}w_{j}u_{j}$ and $H^{(v)}=\sum_{j=1}
^{N}w_{j}v_{j}$. The weights $w_{j}=k_{j}/\left(
  \sum_{\ell=1}^{N} k_{\ell} \right) =k_{j}/k_{total}$ take
into account the difference in contributions of different nodes to the
global mean field, depending on their degrees
(cf.~\cite{Pastor,Ichinomiya}). Thus, the local fields are taken to be
proportional to the degree of a node, ignoring the details of its
actual connections.

With this approximation, the individual activator-inhibitor system on
each node interacts only with the global mean fields, and its dynamics
is described by
\begin{eqnarray}
  \frac{d}{dt} u(t) &=& f(u, v) + \beta (H^{(u)} - u), \cr \cr
  \frac{d}{dt} v(t) &=& g(u, v) + \sigma \beta (H^{(v)} - v).
  \label{Eq:Single}
\end{eqnarray}
We have dropped here the index $i$, since all nodes obey the same
equations, and introduced the parameter $\beta(i) = \varepsilon
k_{i}$. If diffusion ratio $\sigma$ is fixed and the global mean
fields $H^{(u)}$ and $H^{(v)}$ are given, the parameter $\beta$ plays
the role of a bifurcation parameter that controls the dynamics of each
node. Equations~(\ref{Eq:Single}) have a single stable fixed point
when $\beta=0$ (i.e. $\varepsilon=0$), and, as $\beta$ is increased, this
system typically undergoes a saddle-node bifurcation that gives rise
to a new stable fixed point.

\begin{figure}[htbp]
  \begin{center}
    \includegraphics[width=1.0\hsize]{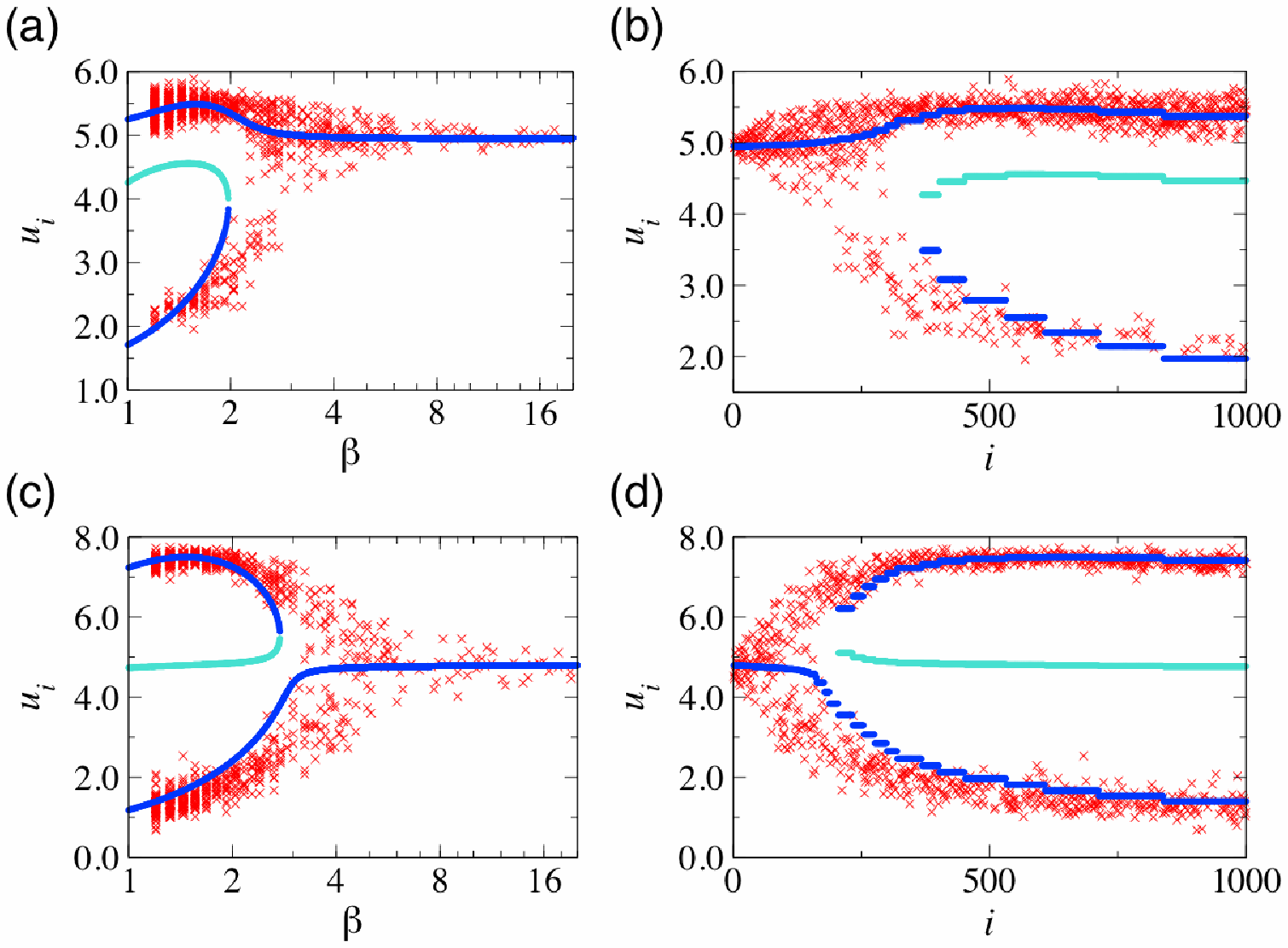}
  \end{center}
  \caption{Developed Turing patterns and mean-field bifurcation
    diagrams.  Stationary Turing patterns compared with the
    bifurcation diagrams of the activator-inhibitor system on a single
    node coupled to global mean fields. The parameters are
    $\varepsilon=0.12$ and (a,b) $\sigma=15.6$, (c,d)
    $\sigma=30$. Blue curves (dots) indicate stable branches and
    light-blue curves (dots) correspond to the unstable
    branches. Crosses show the computed Turing patterns.  The global
    mean fields are $(H^{(u)}, H^{(v)}) = (4.95, 9.97)$ for
    $\sigma=15.6$ and $(H^{(u)}, H^{(v)}) = (4.8, 9.9)$ for
    $\sigma=30$.}
\label{Fig6}
\end{figure}

As an example, we have computed stationary Turing patterns for the
Mimura-Murray model by numerical integration of
equations~(\ref{Eq:RD}) and determined the respective global mean
fields $H^{(u)}$ and $H^{(v)}$ at $\sigma=15.6$ and
$\sigma=30$. Substituting these computed global mean fields into
equations~(\ref{Eq:Single}), bifurcation diagrams for a single node
have been obtained (solid curves in Fig.~\ref{Fig6} (a,c)).  In this
example, one of the two stable branches vanishes by another
saddle-node bifurcation when $\beta$ is increased further.  These
diagrams can be compared with the actual stationary Turing
patterns. Each node $i$ in the network is characterized by its degree
$k_{i}$, so that it possesses a certain value of the bifurcation
parameter, $\beta=\varepsilon k_{i}$. Therefore, the Turing pattern
can be projected onto these bifurcation diagrams, as shown by crosses
in Fig.~\ref{Fig6}(a,c). We see a relatively good agreement between
the stable branches and the data from the actual Turing patterns.
Furthermore, we directly compare in Fig.~\ref{Fig6}(b,d) the computed
Turing patterns with the mean-field predictions, based on
equations~(\ref{Eq:Single}). The Turing patterns are well fitted by
the stable branches, though the scattering of numerical data gets
enhanced near the branching points.

In the Supplementary information, a similar analysis is performed for
the Brusselator model. This model has a different bifurcation diagram
in the presence of external fields. Nonetheless, a good agreement with
the predictions of the mean-field theory is again observed. Thus,
fully developed Turing patterns in the networks are essentially
explained by the bifurcation diagrams of a single node coupled to
global mean fields, with the coupling strength determined by the
degree of the respective network node.  The mean-field theory is
generally not applicable for the localized Turing patterns found below
the Turing instability threshold.

\section{Discussion and conclusions}

The fingerprint property of the classical Turing instability in
continuous media is the spontaneous formation of periodic patterns
above the critical point.  Our investigations of the Turing problem
for large random networks have revealed that, while the bifurcation
remains essentially the same, properties of the emerging patterns are
strongly different.  In the networks, the critical Turing mode is
localized on a subset of network nodes with the degrees close to a
characteristic degree controlled by the mobility of species.  The
final stationary patterns are much different from the critical mode.
Multistability, i.e. coexistence of a number of various stationary
patterns for the same parameter values, is typically found and the
hysteresis phenomena are observed.  Above the instability threshold,
network Turing patterns can be well understood in the framework of the
mean-field approximation.  In this approximation, each network element
is coupled to certain global fields collectively determined by the
entire system, and interactions of the element with its neighbor nodes
are neglected.  The strength of coupling to the global fields depends
on the number of links connecting an element to the rest of the
network.

The Turing instability is also possible in regular lattices
representing a special case of networks.  An activator-inhibitor
system on a lattice can be viewed as a finite-difference approximation
for the respective reaction-diffusion problem in the space and Turing
patterns on the lattices should therefore have almost the same
properties as in the continuous media.  The divergent behavior,
characteristic for large random networks, must be related to a
difference in the structural properties of such systems.  Diameters of
random networks are relatively small and nodes in such networks cannot
be separated by large distances (roughly estimated as $L = \ln N$ for
the Erd\"os-R\'enyi and scale-free networks~\cite{Havlin}).  For
comparison, a cubic lattice with $N$ nodes in the $d$-dimensional
space has the diameter about $L = N^{1/d}$. Thus, a lattice with the
same size $N$ as a random network and a comparable diameter $L$ must
have a high dimension $d \gg 1$.  In lattices with high dimensionality
and short lengths, Turing patterns with many alternating domains are
however not possible and just a few domains (clusters) shall be found,
resembling what is indeed observed by us in large random networks.

Because of their small diameters, diffusional mixing in random
networks is strong.  Large random networks are, therefore,
structurally much closer to the globally coupled systems than to the
low-dimensional lattices.  Globally coupled activator-inhibitor
populations have previously been considered and spontaneous separation
of the elements into two groups has also been found in such
systems~\cite{Mizuguchi}.  There is, however, an important further
aspect distinguishing random networks from the lattices or simple
globally coupled systems.  All nodes in a lattice (or in a globally
coupled population) are equivalent and have the same degree (number of
neighbors).  In contrast to this, random networks effectively
represent strongly heterogeneous systems. They are characterized by
broad degree distributions (less broad but still relatively wide for
the finite-size Erd\"os-R\'enyi networks).  This heterogeneity becomes
essential in the problems involving diffusion.  Under the same
concentration gradients across the links, a node with a higher degree
receives a larger incoming flux from the neighboring nodes.

The heterogeneity is responsible for the localization of Laplacian
eigenvectors on the subsets of nodes with close degrees.  Laplacian
eigenvectors of networks are known to play an important role in the
synchronization phenomena.  However, only the second and the last of
such eigenvectors are significant there (see ~\cite{Boccaletti}). In
contrast, in network Turing problems, the entire Laplacian spectrum
becomes significant.  By varying the diffusional mobility of species,
critical Turing modes corresponding to different Laplacian
eigenvectors are realized.

In the present study, a general framework for the analysis of network
Turing patterns has been proposed. Numerical investigations,
confirming the theory, have been performed for the ecological
predator-prey Mimura-Murray model~\cite{Mimura} and for the classical
chemical Brusselator model~\cite{Prigogine}.  Both models belong to
the activator-inhibitor class.  There is moreover a different broad
class of models where the first species is characterized by
autocatalytic growth (i.e., represents an activator) and it consumes
for its growth the second species which effectively represents a
renewable resource (see, e.g.,~\cite{Mikhailov}).  In these systems,
growth of the activator leads to the depletion of the renewable
resource, which has an inhibitory effect on the activator.  The
considered Turing instabilities should also exist in such other
network-organized systems.

The results of our study may be important in a broad range of
applications.  Turing instabilities can generally be expected in
various cellular, ecological or epidemic networks in Nature and their
detection and observation represent a major scientific challenge. With
the progress in engineering of synthetic ecosystems~\cite{Weber},
artificial ecological networks exhibiting Turing patterns can be
designed in the future.

\section{Methods}

Since $u$ is the activator and $v$ is the inhibitor in
equations~(\ref{Eq:RD}), partial derivatives of $f(u, v)$ and $g(u,
v)$ at $(\bar{u}, \bar{v})$ should satisfy the following conditions:
$f_{u} = \left. \partial f/\partial u\right\vert _{(\bar{u},
  \bar{v})}>0$, $f_{v}=\left.  \partial f/\partial u\right\vert
_{(\bar{u},\bar{v})}<0$, $g_{u}=\left.  \partial g/\partial
  u\right\vert _{(\bar{u},\bar{v})}>0$, and $g_{v}=\left. \partial
  g/\partial v\right\vert _{(\bar{u},\bar{v})}<0$.  The uniform
stationary state of the system $(u_{i}, v_{i}) = (\bar{u}, \bar{v})$
for all $i=1, \cdots, N$ is assumed to be linearly stable in the
absence of diffusion, which requires $f_u + g_v < 0$ and $f_u g_v -
f_v g_u > 0$.

In the Mimura-Murray model~\cite{Mimura}, $u$ and $v$ correspond to
the prey and the predator densities.  In this model, we have $f(u, v)
= \{(a + b u - u^{2}) / c - v \} u$ and $g(u, v) = \{ u - (1 + d v )\}
v$, where the parameters have been chosen as $a = 35$, $b = 16$, $c =
9$, and $d = 2/5$ in the present study, yielding the fixed point
$(\bar{u}, \bar{v}) = (5, 10)$.  In the Brusselator
model~\cite{Prigogine}, variables $u$ and $v$ correspond to densities
of chemical activator and inhibitor species.  Here we have $f(u, v) =
p - (r + 1) u + u^2 v$ and $g(u, v) = r u - u^2 v$ and the parameters
have been chosen as $p = 1$ and $r = 1.8$.  The fixed point is
$(\bar{u}, \bar{v}) = (1, 1.8)$.

Scale-free networks are generated by the preferential attachment
algorithm of Bar\'abasi and Albert~\cite{Barabasi,Barabasi2}, in which
nodes with larger degrees tend to acquire more links.  Starting from
$m$ fully connected initial nodes, we are adding $m$ new connections
at each iteration step, so that the mean degree is $\langle k\rangle
\simeq 2 m$.  The simple Erd\"os-R\'enyi networks are generated by
preparing $N$ nodes and then randomly connecting two arbitrary nodes
with probability $q$, yielding the mean degree of $\langle k \rangle
\simeq N q$~\cite{Barabasi2}.

The Laplacian $L_{ij}$ of any network is a real, symmetric, and
negative semi-definite matrix~\cite{Mohar}.  All eigenvalues are real
and non-positive.  The eigenvectors are orthonormalized as
$\sum_{i=1}^{N} \phi_{i}^{(\alpha)} \phi_{i}^{(\beta)} =
\delta_{\alpha,\beta}$ where $\alpha, \beta = 1, \cdots, N$.

The linear stability analysis is performed in close analogy to the
classical case of continuous media. Small perturbations $\delta u_i$
and $\delta v_i$ obey linearized differential equations
\begin{eqnarray}
  \frac{d}{dt} {\delta u}_{i}(t) &=& f_{u} \delta u_{i} + f_{v} \delta v_{i} + \varepsilon \sum_{j=1}^{N} L_{ij} \delta u_{j}, \cr \cr
  \frac{d}{dt} {\delta v}_{i}(t) &=& g_{u} \delta u_{i} + g_{v} \delta v_{i} + \sigma \varepsilon  \sum_{j=1}^{N} L_{ij} \delta v_{j}.
\end{eqnarray}
Expanding $\delta u_i$ and $\delta v_i$ over the Laplacian normal
modes $\phi_i^{(\alpha)}$ as described in the main text, the following
eigenvalue equation is obtained:
\begin{equation}
  \lambda_{\alpha}
  \left(
    \begin{array}{c}
      1 \cr
      B_\alpha
    \end{array}
  \right)
  =
  \left(
    \begin{array}{cc}
      f_u + \varepsilon \Lambda_{\alpha} & f_v \cr
      g_u & g_v + \sigma \varepsilon \Lambda_{\alpha}
    \end{array}
  \right)
  \left(
    \begin{array}{c}
      1 \cr
      B_\alpha
    \end{array}
  \right).
\end{equation}
From the characteristic equation $\{ \lambda_{\alpha} - f_{u} -
\varepsilon \Lambda_{\alpha} \} \{ \lambda_{\alpha} - g_{v} -\sigma
\varepsilon \Lambda_{\alpha} \} - f_{v} g_{u} = 0$, a pair of conjugate
growth rates are obtained for each Laplacian mode as $
\lambda_{\alpha} = (1/2) \{ f_{u} + g_{v} + ( 1 + \sigma) \varepsilon
\Lambda_{\alpha} \pm [ 4f_{v}g_{u} + \left( f_{u}-g_{v} +(1-\sigma)
  \varepsilon \Lambda_{\alpha} \right)^{2} ]^{1/2} \} $.  Only the upper
branch can become positive and it is always chosen as
$\lambda_{\alpha}$ in our analysis.
From the condition that $\lambda_{\alpha}$ touches the horizontal axis at
its maximum, the critical value of $\sigma$ is determined
as $ \sigma_{c} = \{ f_{u} g_{v} - 2 f_{v} g_{u} +2 [ f_{v} g_{u} (
f_{v} g_{u} - f_{u} g_{v} ) ]^{1/2} \} / f_{u}^{2}$ and the critical
Laplacian eigenvalue is determined for a given $\varepsilon$
as $ \Lambda_c = \{ (f_{u} - g_{v}) \sigma_c - \sqrt{ |f_v| g_u
  \sigma_c } ( \sigma_c + 1 ) \} / \{ \varepsilon \sigma_c ( \sigma_c - 1
) \}$.
The critical eigenvector in the $(u, v)$ plane is given by $(1,
B_{c})$ where $B_{c} = \{ - f_u + g_v + (\sigma_c - 1) \varepsilon
\Lambda_{c} + [ 4 f_v g_u + \left( f_u - g_v - ( \sigma_c - 1 )
  (\varepsilon \Lambda_{c}) \right)^2 ]^{1/2} \} / (2 f_v )$.
These expressions coincide with the respective expressions for the
continuous media (see, e.g.,~\cite{Mikhailov}), if we replace there
$\Lambda$ by $-q^2$, where $q$ is the wavenumber of the plane wave
mode.

\begin{flushleft} 
  {\bf Acknowledgments}
\end{flushleft}

Financial support of the Volkswagen Foundation (Germany) and the MEXT
(Japan, Kakenhi 19762053) is gratefully acknowledged.

  \begin{flushleft}
    {\bf Competing financial interests}
  \end{flushleft}

  The authors declare no competing financial interests.

\newpage

\newpage

\begin{thebibliography}{99}

\bibitem{Turing} Turing, A. M., The chemical basis of morphogenesis,
  {\it Phil. Trans. R. Soc. London B: Biol. Sci.} {\bf 237}, 37
  (1952).

\bibitem{Prigogine} Prigogine, I. and Lefever, R. Symmetry breaking
  instabilities in dissipative systems. II. {\it J. Chem. Phys.} {\bf
    48}, 1695 (1968).

\bibitem{Mimura} Mimura, M. and Murray, J. D., Diffusive prey-predator
  model which exhibits patchiness, {\it J. Theor. Biol.} {\bf 75}, 249
  (1978).

\bibitem{DeKepper} Castets, V., Dulos, E., Boissonade, J. and De
  Kepper, P., Experimental evidence of a sustained standing
  Turing-type nonequilibrium chemical pattern, {\it Phys. Rev. Lett.}
  \textbf{64}, 2953 (1990).

\bibitem{Ouyang} Ouyang, Q. and Swinney, H. L., Transition from a
  uniform state to hexagonal and striped Turing patterns, {\it Nature}
  {\bf 352}, 610 (1991).

\bibitem{Sick} Sick, S., et al., WNT and DKK determine hair follicle
  spacing through a reaction-diffusion mechanism, {\it Science} {\bf
    314}, 1447 (2006).

\bibitem{Othmer} Othmer, H. G., Scriven, L. E., Instability and
  dynamic pattern in cellular networks, {\it J.  Theor. Biol.} {\bf
    32}, 507 (1971).

\bibitem{Hanski} Hanski, I., Metapopulation dynamics, {\it Nature}
  {\bf 396}, 41 (1998).

\bibitem{Urban} Urban, D., and Keitt, T., Landscape connectivity: a
  graph-theoretic perspective, {\it Ecology} {\bf 82}, 1205 (2001).

\bibitem{Hufnagel} Hufnagel, L., Brockmann, D., and Geisel, T.,
  Forecast and control of epidemics in a globalized world, {\it
    Proc. Natl. Acad. Sci. USA} {\bf 101}, 15124 (2004).

\bibitem{Murray} Murray, J. D., {\it Mathematical Biology}, Springer,
  Belin, 2003.

\bibitem{Liu} Liu, Q. -X., and Jin, Z., Formation of spatial patterns
  in epidemic model with constant removal rate of the infectives, {\it
    J. Stat. Mech.} P05002 (2007).

\bibitem{Boccaletti} Boccaletti, S., Latora, V., Moreno, Y., Chavez, M., and Hwang, D.-U., Complex networks: structure and dynamics, {\it Phys. Rep.} {\bf 424}, 175 (2006).

\bibitem{Arenas} Arenas, A., D\'iaz-Guilera, A., Kurths, J., Moreno,
  Y., and Zhou, C., Synchronization in complex networks, {\it
    Phys. Rep.} {\bf 469}, 93 (2008).

\bibitem{Othmer2} Othmer, H. G., Scriven, L. E., Nonlinear aspects of
  dynamic pattern in cellular networks, {\it J. Theor. Biol.} {\bf
    43}, 83 (1974).

\bibitem{Horsthemke} Horsthemke, W., Lam, K., and Moore, P. K., Network
  topology and Turing instability in small arrays of diffusively
  coupled reactors, {\it Phys. Lett. A} {\bf 328}, 444 (2004).

\bibitem{Moore} Moore, P. K., and Horsthemke, W., Localized patterns
  in homogeneous networks of diffusively coupled reactors, {\it
    Physica D} {\bf 206}, 121 (2005).

\bibitem{Barabasi} Barab\'asi, A.-L., and Albert, R., Emergence of
  scaling in random networks, {\it Science} {\bf 286}, 509 (1999).

\bibitem {Barabasi2} Albert, R., and Barab\'asi, A.-L., Statistical
  mechanics of complex networks, {\it Rev. Mod. Phys.} {\bf 74}, 47
  (2002).
  
\bibitem{Mohar} Mohar, B., The Laplacian spectrum of graphs, in {\it
    Graph Theory, Combinatorics and Applications}, Y. Alavi {\it et
    al.}, Eds., New York: Wiley, 1991, pp. 871-898.

\bibitem{Mikhailov} Mikhailov A. S., {\it Foundations of Synergetics
    I. Distributed Active Systems} (2nd revised edition), Springer,
  1994.

\bibitem{Menzinger} McGraw, P. N., and Menzinger, M., Laplacian
  spectra as a diagnostic tool for network structure and dynamics,
  {\it Phys. Rev. E} {\bf 77} 031102 (2008).

\bibitem{Pastor} Pastor-Satorras R., and Vespignani, A., Epidemic
  spreading in scale-free networks, {\it Phys. Rev. Lett.} {\bf 86}
  3200 (2001).

\bibitem{Ichinomiya} Ichinomiya, T., Frequency synchronization in a
  random oscillator network, {\it Phys. Rev. E} {\bf 70} 026116
  (2004).

\bibitem{Havlin} Cohen, R. and Havlin, S., Scale-free networks are
  ultrasmall, {\it Phys. Rev. Lett.} {\bf 90}, 058701 (2003).

\bibitem{Mizuguchi} Mizuguchi, T., and Sano, M., Proportion regulation
  of biological cells in globally coupled nonlinear systems, {\it
    Phys. Rev. Lett.} {\bf 75}, 966 (1995).

\bibitem{Weber} Weber, W., Baba, M. D.-E. and Fussenegger, M.,
  Synthetic ecosystems based on airborne inter- and intrakingdom
  communication, {\it Proc. Natl. Acad. Sci. USA} {\bf 104}, 10435
  (2007).

\end{thebibliography}
\end{document}